\documentclass[letter,preprint]{jpsj3}
\usepackage{txfonts}
\def\vector#1{\mbox{\boldmath $#1$}}
\title{Incommensurate Magnetic Structure in the Cubic Noncentrosymmetric Ternary Compound Pr$_5$Ru$_3$Al$_2$}

\author{Koya Makino$^1$\thanks{makino@mail.tagen.tohoku.ac.jp}, Daisuke Okuyama$^1$, Maxim Avdeev$^2$ and Taku J Sato$^1$}
\inst{$^1$Institute of Multidisciplinary Research for Advanced Materials, Tohoku University,2-1-1 Katahira, Aoba, Miyagi 980-8577, Japan \\
$^2$Bragg Institute, Australian Nuclear Science and Technology Organisation, Locked Bag 2001, Kirrawee DC NSW 2232, Australia} 

\abst{Magnetic susceptibility and neutron powder diffraction experiments have been performed on the noncentrosymmetric ternary compound Pr$_5$Ru$_3$Al$_2$. The previously reported ferromagnetic transition at 24~K was not detected in our improved quality samples. Instead, magnetic ordering was observed in the DC magnetic susceptibility at $T_\textrm{N}\simeq3.8$~K. Neutron powder diffraction further indicates that an incommensurate magnetic structure is established below $T_\textrm{N}$, with the magnetic modulation vector $\vector{q}\simeq(0.066, 0.066, 0.066)$~(r.l.u.). A candidate for the magnetic structure is proposed using the representation analysis.}

\begin{document}
\maketitle

Non-colinear incommensurate magnetic structures, such as a helical magnetic structure, have attracted renewed interests because of the recent discovery of novel topological spin textures, for instance magnetic skyrmion\cite{MnSi_S,FeCoSi,FeGe,Cu2OSeO3,Cu2OSeO3_2,CoZn} and chiral magnetic soliton lattice\cite{CM,CuB2O4,MnSi_CM}. A helical magnetic structure is stabilized by the magnetic frustration\cite{MnO2,Villain} and/or Dzyaloshinsky-Moriya (DM) interaction\cite{Dzyaloshinsky,Moriya}. The latter can be finite for the systems that lack inversion symmetry.
It is well known that the DM interaction uniquely selects helicity of the magnetic structure, and hence can give rise to various intriguing topological quantum effects such as the topological Hall effect in the magnetic skyrmion phase\cite{THE}. Henceforth, many noncentrosymmetric compounds have been synthesized recently.

The ternary compound Pr$_5$Ru$_3$Al$_2$ is one of such noncentrosymmetric compounds\cite{Murashova}. It belongs to the noncentrosymmetric and nonmirrorsymmetric cubic $I2_13$ space group. For this symmetry the finite DM interaction is allowed, and hence an incommensurately modulated magnetic structure is expected.  Nonetheless, the earlier DC magnetic susceptibility measurement only finds the ferromagnetic transition at 24~K\cite{Murashova}. The lowest temperature of the earlier measurement is 4.2~K, and no detailed information is available for the magnetic behavior at lower temperatures.

In this study, we undertook the DC magnetization measurements in the wider temperature range 1.9~K $<$ $T$ $\leq$ 300~K. The sample used in this experiment has greatly improved quality. Using the high quality sample, we found that the ferromagnetic transition at 24~K reported earlier is extrinsic, and indeed Pr$_5$Ru$_3$Al$_2$ shows magnetic ordering at the further lower temperature $T_\textrm{N}\simeq3.8$~K. We also performed neutron powder diffraction and confirmed a formation of the incommensurate magnetic structure below $T_\textrm{N}$ with the very small magnetic modulation vector  $\vector{q}\simeq(0.066, 0.066, 0.066)$~(r.l.u.). This indicates that Pr$_5$Ru$_3$Al$_2$ is a rare example of \emph{4f}-electron-based cubic noncentrosymmetric compounds with the long wavelength magnetic modulation, being closely related to its \textit{3d} analogue MnSi\cite{MnSi}.

\begin{figure}
\begin{center}
\includegraphics*[width=85mm,clip]{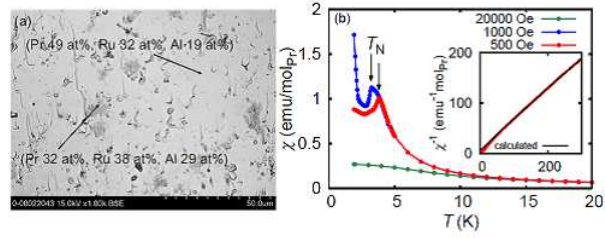}
\end{center}
\caption{(Color online) (a) The BEI obtained using SEM, and composition analysis results by EDX, of the annealed sample. (b) Temperature ($T$) dependence of the DC magnetic susceptibility ($\chi$) for the annealed sample at the magnetic fields $H=500$~Oe, 1000~Oe and 20000~Oe. The inset shows temperature dependence of the inverse susceptibility along with the modified Curie-Weiss law fit (black line).}
\label{f1}
\end{figure}

\begin{figure}
\begin{center}
\includegraphics*[width=85mm,clip]{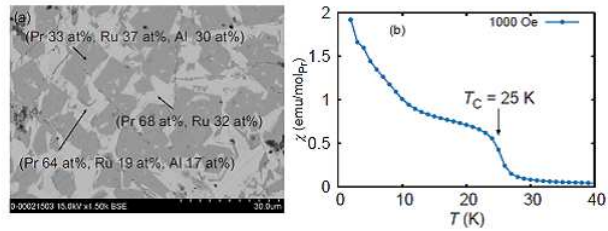}
\end{center}
\caption{(Color online) (a) The BEI and composition analysis results of the as-cast sample. (b) Temperature ($T$) dependence of the magnetic susceptibility ($\chi$) for the as-cast sample at the magnetic field $H=1000$~Oe.}
\label{f2}
\end{figure}

Polycrystalline samples of Pr$_5$Ru$_3$Al$_2$ were synthesized from constituent elements Pr (99.9 at.\%), Ru (99.95 at.\%) and Al (99.99 at.\%) by arc-melting in an Ar atmosphere.\cite{Murashova} The as-cast samples were sealed into quartz tubes with 0.3~atm of Ar gas, and annealed at 650~$\degC$ for 400~h.
Quality of the annealed samples was checked by a SEM (HITACHI, SU6600) with energy dispersive X-ray analysis (EDX).
The DC magnetization was measured between $T$ = 1.9~K and 300~K under the external magnetic fields $H$ = 500~Oe, 1000~Oe and 20000~Oe using a superconducting quantum interference device magnetometer (Quantum Design, MPMS-XL5).The zero-field cooling (ZFC) protocol was used for the magnetization measurement.

Neutron powder diffraction experiment was performed using the high-resolution powder diffractometer ECHIDNA\cite{ECHIDNA} of the Australian Nuclear Science and Technology Organization. Neutrons with the wavelength $\lambda=2.4395$~$\r{A}$ were selected by a monochromator using the Ge 331 reflections. The powdered sample was loaded in a vanadium can, and was set in the standard ILL-type Orange cryostat. Diffraction patterns were taken at 300~K, 10~K, 3~K, and 1.5~K under zero magnetic field. 
The Rietveld refinements were performed for the obtained data using the FULLPROF suite.\cite{fullprof} Visualization of the crystal and magnetic structures was done using the VESTA software.\cite{vesta}

The back scattered electron image (BEI) of the annealed sample together with the composition analysis results by EDX, is shown in {Fig.~\ref{f1}(a)}. According to the EDX result, the main phase of the annealed sample is Pr$_5$Ru$_3$Al$_2$, whereas a small amount of contaminating Pr$_3$Ru$_4$Al$_3$ phase was detected.
Temperature ($T$) dependence of the magnetic susceptibility ($\chi$) for the annealed sample is shown in {Fig.~\ref{f1}(b)}. The susceptibility shows a paramagnetic increase as temperature decreases, which can be seen as the usual linear behavior in the inverse susceptibility shown in the inset of  {Fig.~\ref{f1}(b)}.  The high temperature part of the data was fitted using the modified Curie-Weiss law:
\[\chi(T)=\chi_0+\frac{C}{T-\theta_{\textrm{W}}}\cdot\]  
For the fitted temperature range 100~K $<$ $T$ $<$ 300~K, we obtained the optimum parameters as: $\chi_0=2.4(2)\times10^{-4 }$~emu/mol, $C=1.57(1)$~(emu$\cdot$K)/mol, and $\theta_\textrm{W}=-9.6(4)$~K. The value of the Curie constant $C$ corresponds to the effective magnetic moment 3.5(1)~$\rm{\mu}_{\rm{eff}}$, which is very close to the value of free Pr$^{3+}$ ion (3.58~$\rm{\mu_B}$). This indicates that the magnetic moments are dominantly carried by the Pr$^{3+}$ ions with negligible contributions from the Ru atom. These results are consistent with the earlier study.\cite{Murashova} At the lower temperature $T=24$~K, the earlier study detected ferromagnetic ordering. This ordering was not observed in our annealed sample. Instead, ordering behavior was observed at the further lower temperature 3.8~K under the external field $H = 500$~Oe. This ordering is strongly magnetic field dependent, and was observed at 3.25~K under $H = 1000$~Oe. 

To elucidate the origin of the ferromagnetic ordering reported earlier, we performed similar measurements on the as-cast sample. The BEI of the as-cast sample is shown in {Fig.~\ref{f2}(a)}, together with the composition analysis results. The composition analysis shows that this as-cast sample does not contain Pr$_5$Ru$_3$Al$_2$ but contains Pr$_3$Ru, Pr$_4$RuAl, and Pr$_3$Ru$_4$Al$_3$. 
Temperature  dependence of  the DC magnetic susceptibility  for the as-cast sample is shown  in {Fig.~\ref{f2}(b)}. A ferromagnetic transition was observed at 25~K, which is similar to the one observed in the earlier study.
From these results, we conclude that the ferromagnetic transition observed at 24~K in the earlier study originates from one of the three compounds  Pr$_3$Ru, Pr$_4$RuAl, and Pr$_3$Ru$_4$Al$_3$, and hence is not intrinsic property of Pr$_5$Ru$_3$Al$_2$.

\begin{table}
\caption{Refined atomic positions at 300 K and 10 K. The space group is \textit{I}2$_1$3. 58 reflections were used in the refinement with 9 adjustable parameters. For $T=300$~K, the lattice constant, the cell volume, and the calculated density of the cell are $a=9.79542(4)$~\r{A}, $V=939.874(7)$ ~\r{A}$^3$, and $\rho=7.1596$~g/cm$^3$. The R-factors are: \textit{R$_{\emph{p}}$}~=~18.9~\%, \textit{R$_{\emph{wp}}$}~=~15.5~\%, \textit{R$_{\emph{expt}}$}~=~5.9~\%, and $\chi^2$~=~7.1. For $T=10$ K, $a=9.77736(3)$~\r{A}, $V=934.683(6)$~\r{A}$^3$, and $\rho=7.1994$~g/cm$^3$. The R-factors are: \textit{R$_{\emph{p}}$}~=~16.2~\%, \textit{R$_{\emph{wp}}$}~=~13.7~\%, \textit{R$_{\emph{expt}}$}~=~3.8~\%, and $\chi^2$~=~13.4.}
\label{t1}
\begin{center}
\begin{tabular}{llllll}
\hline
\multicolumn{1}{c}{Atom} & \multicolumn{1}{c}{Site} &\multicolumn{1}{c}{$x$}& \multicolumn{1}{c}{$y$}& \multicolumn{1}{c}{$z$}& \multicolumn{1}{c}{$U_{\textrm{iso}}$} \\
\hline
\textit{T}=300 K \\
Pr1 & 12b &0.4395(5) &0 &1/4 &0.017(2) \\
Pr2 & 8a &0.1098(6) &0.1098(6) &0.1098(6) &0.001(3) \\
Ru & 12b &0.8718(5) &0 &1/4 &0.0042(8) \\
Al & 8a &0.2936(6) &0.2936(6) &0.2936(6) &0.010(4) \\
\textit{T}=10 K \\
Pr1 & 12b &0.4393(5) & 0& 1/4&0.0078(18) \\
Pr2 & 8a &0.1087(5) &0.1087(5) &0.1087(5) &0.002(2) \\
Ru & 12b &0.8696(4) &0 &1/4 &0.0046(7) \\
Al & 8a & 0.2935(5) & 0.2935(5) & 0.2935(5) &0.005(3) \\
\hline
\end{tabular}
\end{center}
\end{table}

\begin{figure}
\begin{center}
\includegraphics*[width=80mm,clip]{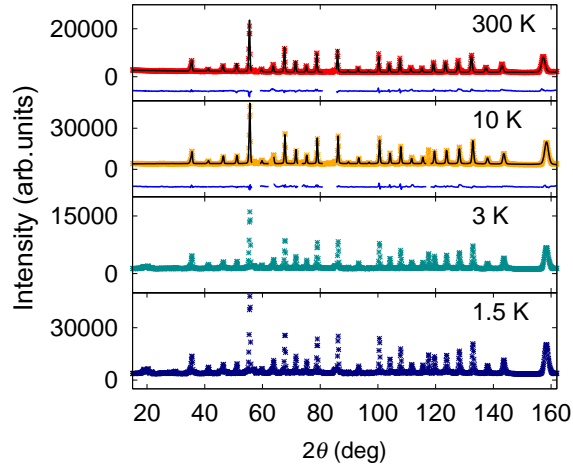}
\end{center}
\caption{(Color online) Powder neutron diffraction patterns of Pr$_5$Ru$_3$Al$_2$ at 300~K, 10~K, 3~K and 1.5~K, together with Rietveld analysis results at 300~K and 10~K. The bottom blue lines show the difference between the observed (cross mark) and calculated (black line) intensities. We excluded the 2$\theta$ regions:, 56.5-59.0 and 81.0-85.5, from Rietveld refinements where the unknown impurity peaks exist. For the data at low temperatures (10~K, 3~K, 1.5~K), we additionally excluded the 2$\theta$ regions:, 62.2-63.8, 72.5-73.5, and 116.8-118.2, from fitting where extra reflections from the Al cryostat windows were observed.}
\label{f3}
\end{figure}
\begin{table}
\caption{Basis vectors (BVs) of irreducible representations (IRs) for the space group $I2_13$ with the magnetic modulation vector $\vector{q}=(\delta, \delta, \delta)$, $\delta\simeq0.066$. Atom numbers in each orbit are defined as: atom 1: ($x_1$,0,1/4), atom 2: (1/4,$x_1$,0), atom 3:(0,1/4,$x_1$) for orbit 1, atom 1: (1/2-$x_1$,0,3/4), atom 2: (3/4,1/2-$x_1$,0), atom 3:(0,3/4,1/2-$x_1$) for orbit 2, atom 1:($x_2$,$x_2$,$x_2$) for orbit 3, or atom 1: (1/2-$x_2$,1-$x_2$,1/2+$x_2$), atom 2: (1/2+$x_2$,1/2-$x_2$,1-$x_2$), atom 3:(1-$x_2$,1/2+$x_2$,1/2-$x_2$) for orbit  4, where $x_1=0.43866$, $x_2=0.10893$. It should be noted that the centering translation +(1/2,1/2,1/2) exists.}
\label{t3}
\begin{center}
\begin{tabular}{ccccc}
\hline
\multicolumn{5}{c}{Orbit 1, Orbit 2 and Orbit 4}\\
		\multicolumn{1}{c}{IR} & \multicolumn{1}{c}{BV} &\multicolumn{1}{c}{Atom 1}& \multicolumn{1}{c}{Atom 2}& \multicolumn{1}{c}{Atom 3}\\
\hline
$\Gamma_1$&$\psi_1$&(1,0,0)&$(0,1,0)$ &$(0,0,1)$\\
{}&$\psi_2$&(0,1,0)&$(0,0,1)$ &$(1,0,0)$\\
{}&$\psi_3$&(0,0,1)&$(1,0,0)$ &$(0,1,0)$\\
$\Gamma_2$&$\psi_4$&(1,0,0)&$(0,-\frac{1}{2}-\frac{\sqrt{3}}{2}i,0)$ &$(0,0,-\frac{1}{2}+\frac{\sqrt{3}}{2}i)$\\
{}&$\psi_5$&(0,1,0)&$(0,0,-\frac{1}{2}-\frac{\sqrt{3}}{2}i)$ &$(-\frac{1}{2}+\frac{\sqrt{3}}{2}i,0,0)$\\
{}&$\psi_6$&(0,0,1)&$(-\frac{1}{2}-\frac{\sqrt{3}}{2}i,0,0)$ &$(0,-\frac{1}{2}+\frac{\sqrt{3}}{2}i,0)$\\
$\Gamma_3$&$\psi_7$&(1,0,0)&$(0,-\frac{1}{2}+\frac{\sqrt{3}}{2}i,0)$ &$(0,0,-\frac{1}{2}-\frac{\sqrt{3}}{2}i)$\\
{}&$\psi_8$&(0,1,0)&$(0,0,-\frac{1}{2}+\frac{\sqrt{3}}{2}i)$ &$(-\frac{1}{2}-\frac{\sqrt{3}}{2}i,0,0)$\\
{}&$\psi_9$&(0,0,1)&$(-\frac{1}{2}+\frac{\sqrt{3}}{2}i,0,0)$ &$(0,-\frac{1}{2}-\frac{\sqrt{3}}{2}i,0)$\\
\hline
\end{tabular}
\end{center}

\begin{center}
\begin{tabular}{ccc}
\hline
\multicolumn{3}{c}{Orbit 3}\\
		\multicolumn{1}{c}{IR} & \multicolumn{1}{c}{BV} &\multicolumn{1}{c}{Atom 1}\\
\hline
$\Gamma_1$&$\psi_1$&$(1,1,1)$ \\
$\Gamma_2$&$\psi_4$&$(1,-\frac{1}{2}-\frac{\sqrt{3}}{2}i,-\frac{1}{2}+\frac{\sqrt{3}}{2}i)$\\
$\Gamma_3$&$\psi_7$&$(1,-\frac{1}{2}+\frac{\sqrt{3}}{2}i,-\frac{1}{2}-\frac{\sqrt{3}}{2}i)$ \\
\hline
\end{tabular}
\end{center}
\end{table}

To study the microscopic spin structure in the low-temperature-ordered phase of Pr$_5$Ru$_3$Al$_2$, we performed neutron powder diffraction measurements at 300~K, 10~K, 3~K and 1.5~K using the annealed sample. The resulting diffraction patterns are shown in {Fig.~\ref{f3}(a)}. For the patterns observed at the paramagnetic temperatures, we performed the Rietveld refinements using the existing structural model with the space group $I2_13$. The result is also shown in {Fig.~\ref{f3}(a)}. Reasonable agreements were seen between the observed and calculated patterns, as evidenced by the difference shown in panels of the data at 300~K and 10~K, indicating the correctness of the existing structure model. Unknown impurity peaks were found in the diffraction patterns, however the strongest impurity peak was 20 times smaller than that of the main phase. 
In this experiment, we used the relatively long wavelength $\lambda=2.4395$ $\r{A}$ to study magnetic reflections in the low-$Q$ region in detail. Hence, we could not obtain reflections in the high-$Q$ region. Therefore, the isotropic atomic displacement parameter {$U_{\textrm{iso}}$ was not reliably determined in the present Rietveld analysis. {Table~\ref{t1}} summarizes refined crystallographic parameters.

At further lower temperatures, magnetic reflections start to develop, and become clearly visible at 3~K and 1.5~K. Inset of {Fig.~\ref{f4}(a)} shows the magnified view of the diffraction patterns in the lower $2\theta$ range measured at 10~K and 1.5~K. Clearly, the 1.5~K data show extra reflections which were not observed at 10~K. The results are consistent with the DC magnetic susceptibility measurement where the transition was observed at $T_\textrm{N}\simeq3.8$~K at low external magnetic field.
Magnetic satellite reflections were observed in the vicinity of nuclear reflections. This clearly indicates that magnetic ordering is neither simple ferromagnetic nor antiferromagnetic, but incommensurately modulated with a long modulation period. Under the cubic symmetry, the eight satellite reflections at $(\pm\delta, \pm\delta, \pm\delta)$ around the 110 nuclear peak appear as three reflections at $Q_1=|(1-\delta, 1-\delta, \pm\delta)|$, $Q_2=|(1\pm\delta, 1\mp\delta, \pm\,or\mp\delta)|$ and $Q_3=|(1+\delta, 1+\delta, \pm\delta)|$ in the powder diffraction pattern. Indeed, this is the case for the positions and the number of magnetic satellite reflections visible in the present powder pattern. Thus, the magnetic propagation vector $\vector{q}$ is most likely along the (1,1,1) direction, \textit{i. e.}, $\vector{q}=(\delta, \delta, \delta)$. From the peak position, we estimate $\delta\simeq0.066$~(r.l.u.). This corresponds  to the modulation with the length of $\sim90$ $\r{A}$. We, therefore, conclude that a very interesting long wavelength incommensurate magnetic structure is formed in this Pr compound.  

\begin{figure}
\begin{center}
\includegraphics*[width=70mm,clip]{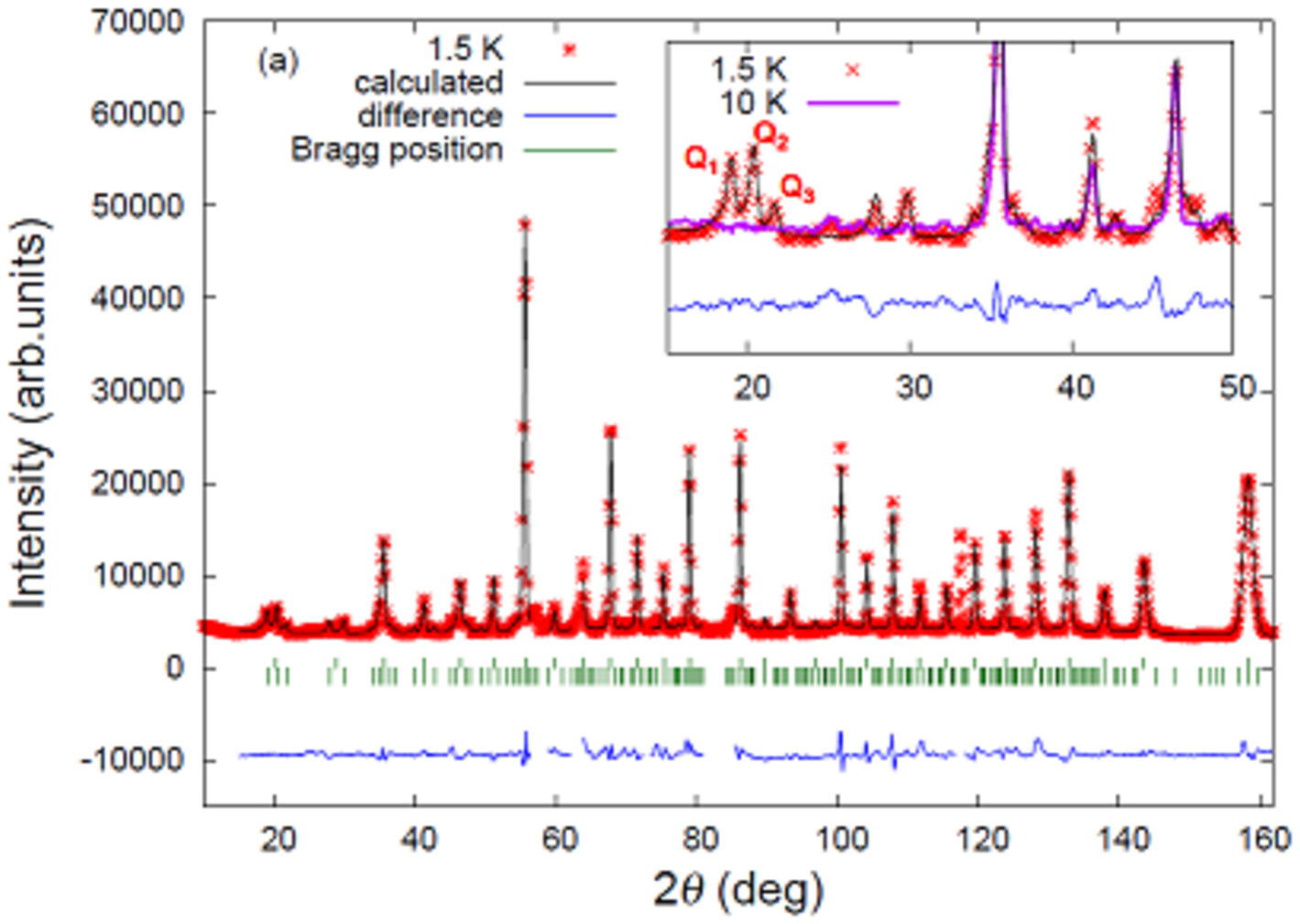}
\includegraphics*[width=80mm,clip]{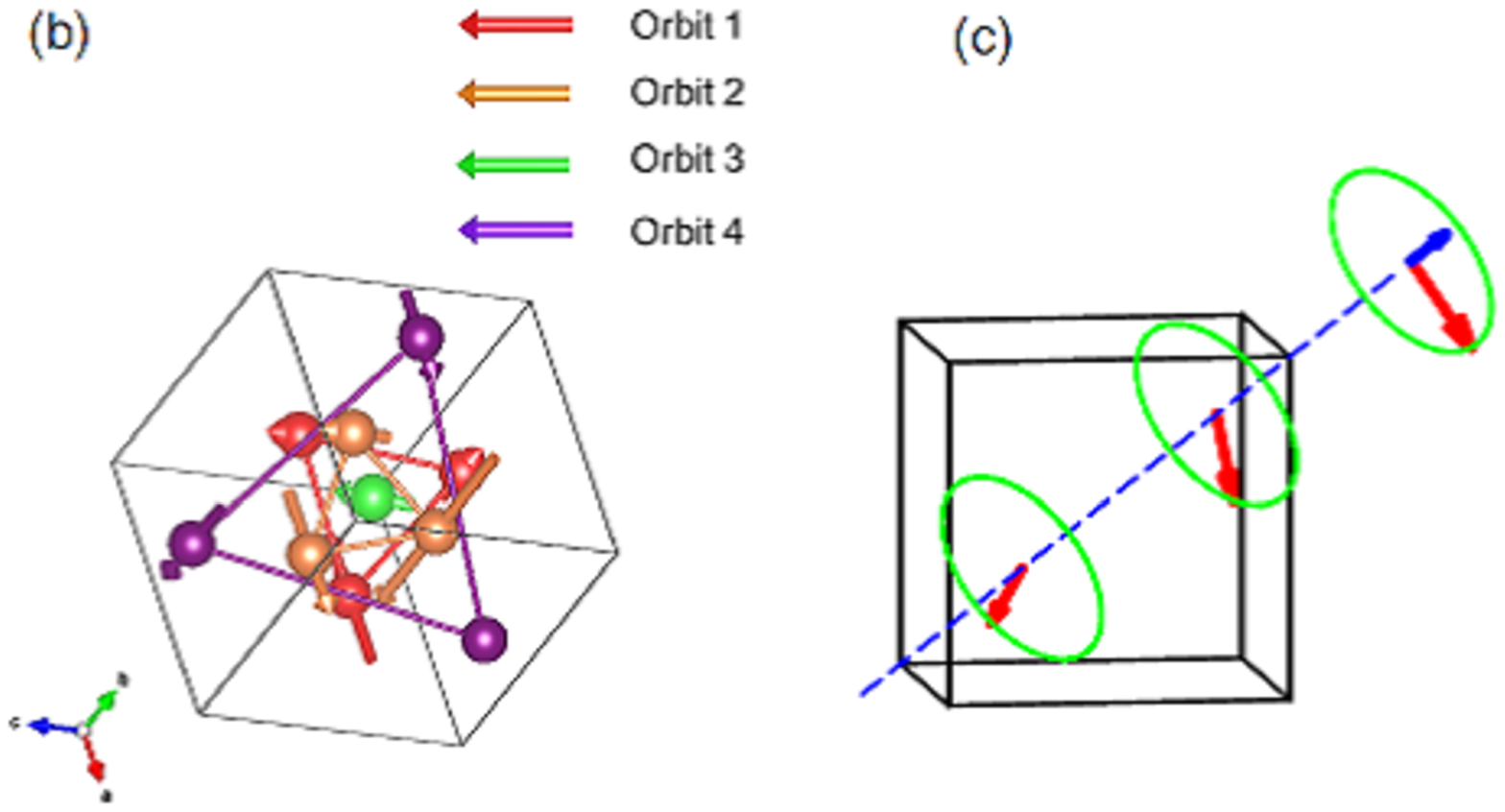}
\end{center}
\caption{(Color online) (a)Rietveld analysis result for $T$ = 1.5~K data shown in Fig.~\ref{f3}. The calculated positions of nuclear and magnetic reflections are indicated by the green ticks. The bottom line represents the difference. The inset displays the magnified view for  $2\theta<50^\circ$. (b) The schematic illustration of the magnetic structure proposed in the present study. For simplicity, centering translation is omitted and only a half portion of the unit cell is illustrated. (c) Spatial modulation of the total Pr$^{3+}$ moment in the orbit 1 for the presently proposed $\Gamma_2$ IR.
}
\label{f4}
\end{figure}

To find a possible microscopic model for the magnetic structure, the magnetic representation analysis\cite{Bertaut,Izyumov,Izyumov2} was performed using the paramagnetic space group $I2_13$ and determined magnetic wave vector $\vector{q}=(\delta, \delta, \delta)$.
For the symmetry operations of the k-subgroup with $\vector{q}=(\delta, \delta, \delta)$, the crystallographically equivalent Pr1(12b) sites become inequivalent, and two orbits (orbit 1 and orbit 2) are formed. Similarly, the Pr2(8a) sites split in the two orbits (orbit 3 and orbit 4). The orbits 1, 2, and 4 contain three atoms, which are symmetrically related by the three fold rotation along the [111] axis. The orbit 3 has only one atom. The basis vectors (BVs) of the irreducible representations (IRs) for the k-subgroup were calculated, and listed in {Table~\ref{t3}}. All the IRs  are one dimensional. Landau theory for the second order phase transition requires a single IR to be selected in the ordered phase, and hence linear combinations of BVs belonging to the same IR have to be taken into account in principle. In addition, BVs for the different orbits can be selected independently. Therefore, the initial model structure should be linear combinations of the BVs in a IR, with independent phases (and amplitudes) for different orbits. However, the number of the magnetic reflections observed experimentally is not enough to refine all the parameters in this general model. Therefore, we further assume that a single BV is selected for each orbit for the sake of simplicity. Using this assumption, a trial function for the spin structure analysis may be written as:
\[\vector{S}(l, j, i) = \frac{C_i}{2}\left\{\vector\psi_{n(i)}^je^{-i\left[\vector{q}\cdot\left(\vector{R}_j+\vector{R}_l\right)+2\pi\phi_i\right]}+\textrm{c.c.}\right\},\]
where $\vector{S}(l, j, i)$ stands for the magnetic moment at the $j$-th atom of the $i$-th orbit in the $l$-th unit cell. $\psi_{n(i)}^j$ is the $n$-th BV for the $j$-th atom of the $i$-th orbit, and given in Table \ref{t3}, whereas, $C_i$ and $\phi_i$ are the amplitude and phase for the BVs for the atoms in the $i$-th orbit. c.c. stands for complex conjugate. Since we assume that the magnetic structure belongs to a single IR, we select all the four BVs, \{${\vector\psi_{n(1)}, \vector\psi_{n(2)}, \vector\psi_{n(3)}, \vector\psi_{n(4)}}$\}, from those belonging to a single IR. For all the combinations of \{${\vector\psi_{n(1)}, \vector\psi_{n(2)}, \vector\psi_{n(3)}, \vector\psi_{n(4)}}$\}, we refined $C_{i=\rm{1,\,2,\,3,\,and\,4}}$ and $\phi_{i=\rm{2,\,3,\,and\,4}}$ using the Rietveld refinement technique. (For the incommensurate structure, only relative values of phases are determined. Thus, we fixed $\phi_1=0$ and refined other phases.) As a result, the best fit was obtained for the $\Gamma_2$ IR, with the BVs being \{$\psi_4$, $\psi_5$, $\psi_4$, $\psi_6$\}, \textit{i. e.} , $n(1)=4$, $n(2)=5$, $n(3)=4$, and $n(4)=6$. The optimum amplitudes and phases are accordingly $C_1=-2.5(3)$~$\rm{\mu_B}$, $C_2=3.8(3)$~$\rm{\mu_B}$, $C_3=-1.0(2)$~$\rm{\mu_B}$, $C_4=2.0(3)$~$\rm{\mu_B}$, $\phi_2=-0.53(2)$, $\phi_3=-0.16(3)$, and $\phi_4=-0.24(3)$. The result of the refinement is shown in {Fig.~\ref{f4}(a)}. The $R$-factor for this result is  $R_{\rm{p}}\sim17.3$~\%, whereas the magnetic $R$-factor $R_{\rm{mag}}\sim31.5$~\%. It may be noted that the obtained amplitudes $C_i$ do not exceed the effective magnetic moment of free Pr$^{3+}$ ion 3.58~$\rm{\mu_B}$. This structure is depicted in Fig.~\ref{f4}(b) for a half of the unit cell.

The total magnetic moment for each orbit, defined as:
\[\vector{m}_i(l) = \frac{C_i}{2}\left\{\sum_{j\in \rm{orbit}\,i}\vector\psi_{n(i)}^je^{-i\left[\vector{q}\cdot\left(\vector{R}_j+\vector{R}_l\right)+2\pi\phi_i\right]}+\textrm{c.c.}\right\},\]
can be calculated as: 
\[\vector{m}_{i=1,\,3}(l)= C_i\Bigl[\cos{X_i(l)},\,\cos(X_i(l)-\textrm{A}),\,\cos(X_i(l)+\textrm{A})\Bigr],\] 
\[\vector{m}_2(l) = C_2\Bigl[\cos(X_2(l)+\textrm{A}),\,\cos{X_2(l)},\,\cos(X_2(l)-\textrm{A})\Bigr],\]
\[\vector{m}_4(l) = C_4\Bigl[\cos(X_4(l)-\textrm{A}),\,\cos(X_4(l)+\textrm{A}),\,\cos{X_4(l)}\Bigr],\]
where $X_i(l)=\vector{q}\cdot\left(\vector{R}_{l}+\vector{R}_{j\in \rm{orbit}\,i}\right)+2\pi\phi_i$, $qR_i=$, and $\textrm{A}=2\pi/3$.
It can be seen from the above results that the size of the total magnetic moment is constant, and that its direction is always perpendicular to $\vector{q}$, \textit{i. e.},  $\vector{m}_i\bot\vector{q}$. Hence, the proposed magnetic structure can be seen as a variation of a helical spin structure by regarding each orbit as a unit block of the magnetic structure. The spatial dependence of the total magnetic moment $\vector{m}_1$ is shown in Fig.\ref{f4}(c). Also the other $\vector{m}_i$'s rotate in the same direction. Here, we refer to this magnetic structure as the block helical structure. It should, nevertheless, be noted that this block helical nature of the orbit 1, 2, and 4 originates from the selection of the single BV for each orbit, and hence mixing of other BVs will degrade the ideal helical rotation of the total magnetic moment. We think that the selection of the single BV is  a reasonable assumption as a first approximation, nonetheless, further single crystal neutron diffraction is apparently necessary to check this assumption. It should be further noted that, although the total moment exhibits helical rotation, the size of single Pr magnetic moments modulates sinusoidally as $|\cos(X_i(l)+\textrm{B})|$ where $\textrm{B}=0,\,2\pi/3,\,\textrm{or}\,-2\pi/3$ except for the orbit 3. Hence, magnetic moments may be vanishingly small on some sites. These sites may be thermodynamically unstable at lower temperatures, and the system may show a further phase transition to a commensurate structure. Lower temperature magnetization study below 2 K is desired to check this possibility. For the orbit 3 there is only one atom on the [111] axis, and the magnetic moment is helically modulated, \textit{i. e.}, $\vector{S}(l, 1, 3)=\vector{m}_3(l)$.

The  incommensurate magnetic structure is not rare in the rare-earth magnetism, such as those found in Tb, Dy, Ho, Er, and Tm\cite{REM}. Nonetheless, those incommensurate structures are stabilized by the frustrating long-range Ruderman-Kittel-Kasuya-Yosida (RKKY) interactions. The incommensurate magnetic structure stabilized by the DM interaction is not widely known in rare-earth compounds. Indeed, we are only aware of the following two examples: YbNi$_3$Al$_9$ and  Yb(Ni$_{0.94}$Cu$_{0.06}$)$_3$Al$_9$. They were reported as \textit{4f}-electron-based chiral magnets with the competing RKKY and DM interactions\cite{YbNiAl}. They belong to the noncentrosymmetric and nonmirrorsymmetric $R32$ space group. The magnetic modulation vector of YbNi$_3$Al$_9$ was determined to be $\vector{q}=(0, 0, 0.8)$ using neutron diffraction.
In contrast to those uni-axial compounds, the present Pr$_5$Ru$_3$Al$_2$ has the much longer modulation period, being closer to the ferromagnetic order. Together with its crystallographic cubic structure, Pr$_5$Ru$_3$Al$_2$ is much closely related to its \textit{3d} analogue MnSi\cite{MnSi}.

In summary, we have performed the DC magnetization and neutron powder diffraction measurements in the ternary compound Pr$_5$Ru$_3$Al$_2$. The ferromagnetic transition reported earlier was attributed to the impurity phases, and instead we found the new anomaly in the DC magnetization measurement at the lower temperature  $T_\textrm{N}\simeq3.8$~K. Our neutron powder diffraction experiment confirmed that an incommensurate magnetic structure is established below $T_\textrm{N}$ with the long modulation period characterized by the modulation vector $\vector{q}=(\delta, \delta, \delta)$, where $\delta\simeq0.066$~(r.l.u.). Using the magnetic representation analysis, we propose a candidate for the incommensurate magnetic structure. This is a rare example of long period modulated structures in the cubic rare-earth compounds.
\begin{acknowledgment}

The authors thank Yusuke Nambu, Yoichi Ikeda, Takahiro Onimaru, Yusuke Tokunaga, Kunihiko Yamauchi, and Tamio Oguchi for stimulating discussions. The authors also thank  Satoshi Ohhashi and Masao Ishi for the technical assistance in SEM measurements. This work was partly supported by Grants-In-Aid for Scientific Research (24224009 and 23244068) from MEXT of Japan, and by Nano-Macro Materials, Devices and System Research Alliance.  The travel expenses for the neutron scattering experiment at ANSTO were partly supported by General User Program for Neutron Scattering Experiments, Institute for Solid State Physics, University of Tokyo.

\end{acknowledgment}

\bibliographystyle{jpsj}
\bibliography{readcube_export}

\end{document}